\documentclass[10pt,conference]{IEEEtran}
\IEEEoverridecommandlockouts
\usepackage{cite}
\usepackage{amsmath,amssymb,amsfonts}
\usepackage{algorithm}
\usepackage{algpseudocode}
\usepackage{algorithmicx}
\usepackage{graphicx}
\usepackage{textcomp}
\usepackage{xcolor}
\usepackage{booktabs}
\usepackage{epstopdf}
\usepackage[T1]{fontenc}
\usepackage{subfigure}
\usepackage{comment}
\usepackage{multirow} 
\usepackage{geometry}
\begin{document}

\title{Adaptive Offloading and Enhancement for Low-Light Video Analytics on Mobile Devices}

\author{\IEEEauthorblockN{Yuanyi~He\IEEEauthorrefmark{1},~Peng~Yang\IEEEauthorrefmark{1},~Tian~Qin\IEEEauthorrefmark{1},~Jiawei~Hou\IEEEauthorrefmark{1},~and~Ning~Zhang\IEEEauthorrefmark{2}}
\IEEEauthorblockA{\IEEEauthorrefmark{1}School of Electronic Information and Communications, Huazhong University of Science and Technology, Wuhan, China \\
   \IEEEauthorrefmark{2}Department of Electrical and Computer Engineering, University of Windsor, Windsor, ON, Canada\\
    \hbox{Email: \IEEEauthorrefmark{1}\{yuanyi$\_$he, yangpeng, qin$\_$tian, jerry$\_$hou\}@hust.edu.cn, \IEEEauthorrefmark{2}ning.zhang@uwindsor.ca}}
}

\maketitle

\begin{abstract}
In this paper, we explore adaptive offloading and enhancement strategies for video analytics tasks on computing-constrained mobile devices in low-light conditions. We observe that the accuracy of low-light video analytics varies from different enhancement algorithms. The root cause could be the disparities in the effectiveness of enhancement algorithms for feature extraction in analytic models. Specifically, the difference in class activation maps (CAMs) between enhanced and low-light frames demonstrates a positive correlation with video analytics accuracy. Motivated by such observations, a novel enhancement quality assessment method is proposed on CAMs to evaluate the effectiveness of different enhancement algorithms for low-light videos. Then, we design a multi-edge system, which adaptively offloads and enhances low-light video analytics tasks from mobile devices. To achieve the trade-off between the enhancement quality and the latency for all system-served mobile devices, we propose a genetic-based scheduling algorithm, which can find a near-optimal solution in a reasonable time to meet the latency requirement. Thereby, the offloading strategies and the enhancement algorithms are properly selected under the condition of limited end-edge bandwidth and edge computation resources. Simulation experiments demonstrate the superiority of the proposed system, improving accuracy up to 20.83\% compared to existing benchmarks.
\end{abstract}

\section{Introduction}
    The advancement of computer vision technology has become indispensable across various domains of Internet of Things (IoT) applications\cite{8057318, yang2023adaptive}. Key to this progress is advanced high-level visual models, such as YOLO, Mask R-CNN, and ResNet, which utilize deep neural networks (DNNs) for video analytics\cite{Kong,zhangOwlPreandPostprocessing2023}. Video analytics can help analyze and interpret videos obtained from visual sensors to accomplish various tasks, such as object detection, semantic segmentation, and object tracking\cite{du2020server}. To promote video analytics on computing-constrained mobile devices, prevalent researches suggest that video frames captured by mobile devices can be transmitted to nearby edge servers for acquiring additional computation resources through wireless networks\cite{10000829,10.1145/3503161.3548033,chengyan,9155524}.

    The establishment of edge clusters with edge servers allows for efficient processing originating from mobile devices\cite{li2023tapfinger, chen2024tilesr,Axiomvision}. The bandwidth resources of the wireless networks for transmission from end devices to edge servers are constrained. And the computation capacity of the edge servers presents diversity\cite{yangpeng}. This creates scheduling opportunities by assigning tasks from end devices to suitable edge servers, which ensures a balance between bandwidth and computation resources throughout the cluster\cite{pan2023joint}.

    In low-light conditions, videos result in reduced visibility, loss of detail, and distortion of colors. Thus, video analytics in low-light conditions suffers more difficulties than that in normal-light conditions\cite{he2023globecom}. Extra computation workloads are suggested to enhance these low-light frames\cite{zhangOwlPreandPostprocessing2023}. Various low-light enhancement algorithms have been proposed, including theory-based algorithms such as histogram equalization (HE) for distribution mapping \cite{coltuc2006exact} and LIME based on Retinex theory \cite{7782813}, as well as learning-based neural network algorithms such as Zero-DCE \cite{guo2020zero} and EnlightenGAN \cite{jiang2021enlightengan}. They are applicable as pre-processing for inference.

    However, the accuracy of low-light video analytics with enhancement algorithms is not consistent with human eye visual quality\cite{he2023globecom}. Existing video quality evaluation metrics, i.e., VMAF, PSMR, and SSIM, are not applicable to evaluate the effectiveness of various low-light video enhancement algorithms. Although video analytics accuracy serves as the direct reflection of enhancement algorithm effectiveness\cite{wu2022edge}, the calculation of real-time low-light video analytics accuracy faces two challenges. Firstly, in practice, the absence of video analytics ground truth hinders the determination of intermediate values that are used for accuracy calculation, e.g., intersection-over-union. Secondly, the impact of various enhancement algorithms on accuracy is unknown in advance. Even on labeled low-light video datasets, edge servers are required to execute all enhancement algorithms to determine the accuracy of the enhanced video for assessing the effectiveness of enhancements, which can be time-consuming. Therefore, it is necessary to design an assessment method to indicate the accuracy of low-light video analytics. Class Activation Maps (CAMs)\cite{zhou2016learning} provide insight into the regions of interest identified by high-level visual models during decision-making processes. Consequently, it offers an opportunity to proactively assess the efficacy of various enhancement techniques.


  This paper introduces a system designed for low-light video analytics across end devices and edge servers. A collection of heterogeneous edge servers forms an edge cluster, where an edge server is the schedule controller. It is responsible for the enhancement quality assessment based on CAMs, and making schedule decisions accordingly. It is essential to select suitable enhancement algorithms and dedicated edge servers to offload for each end device. Given the constraints of limited bandwidth resources between end devices and edges, as well as the computation resources at the edges, an optimization problem is formulated. The goal is to address the trade-off between enhancement quality and latency across all end devices. To solve the problem, a genetic-based scheduling algorithm is proposed. Our main contributions are summarized as follows.
\begin{itemize}
	\item We investigate the analytics accuracy of different low-light videos processed by different enhancement algorithms. Comparing the attributes of their CAMs, we develop a novel enhancement quality assessment to evaluate the effectiveness of various enhancement algorithms.
	\item We design a multi-edge system that achieves adaption to offloading and enhancement for low-light video analytics on mobile devices through wireless networks, to balance the trade-off between enhancement quality and latency. We introduce a genetic-based scheduling algorithm to implement the adaption with limited bandwidth and computation resources.
	\item Our approach exhibits advantages over other baseline methods, with an improvement of accuracy up to 20.83\%. There is only a latency increase of up to 72.30 ms compared to existing algorithms without enhancement.
\end{itemize}

The remainder of this paper is as follows. In Section II, we outline our motivation. Section III elaborates on the system model, problem formulation, and proposed algorithm. Section IV provides an evaluation of the algorithm. Finally, Section V is the conclusion of the paper.

\section{Motivation}
	\subsection{Detection of Different Low-light Videos}
        \label{Detection of different low-light videos}
\begin{figure}[t]
		\centering
		\subfigure[Video 1]{
			\begin{minipage}[b]{0.44\linewidth}
				\includegraphics[width=1\textwidth]{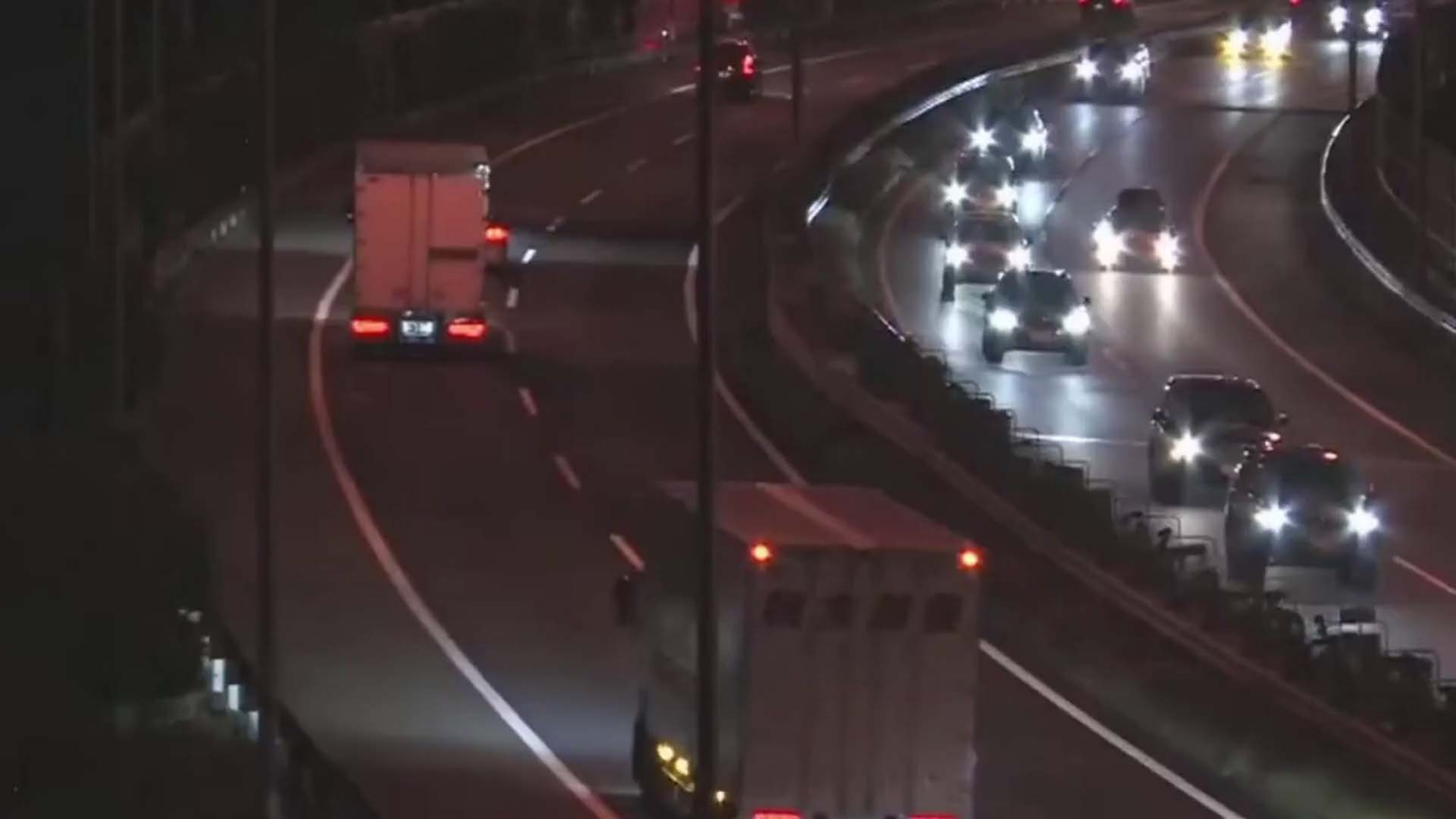} 
			\end{minipage}
			\vspace{-5mm}
			\label{Video 1}
		}
		\subfigure[Video 2]{
			\begin{minipage}[b]{0.425\linewidth}
				\includegraphics[width=1\textwidth]{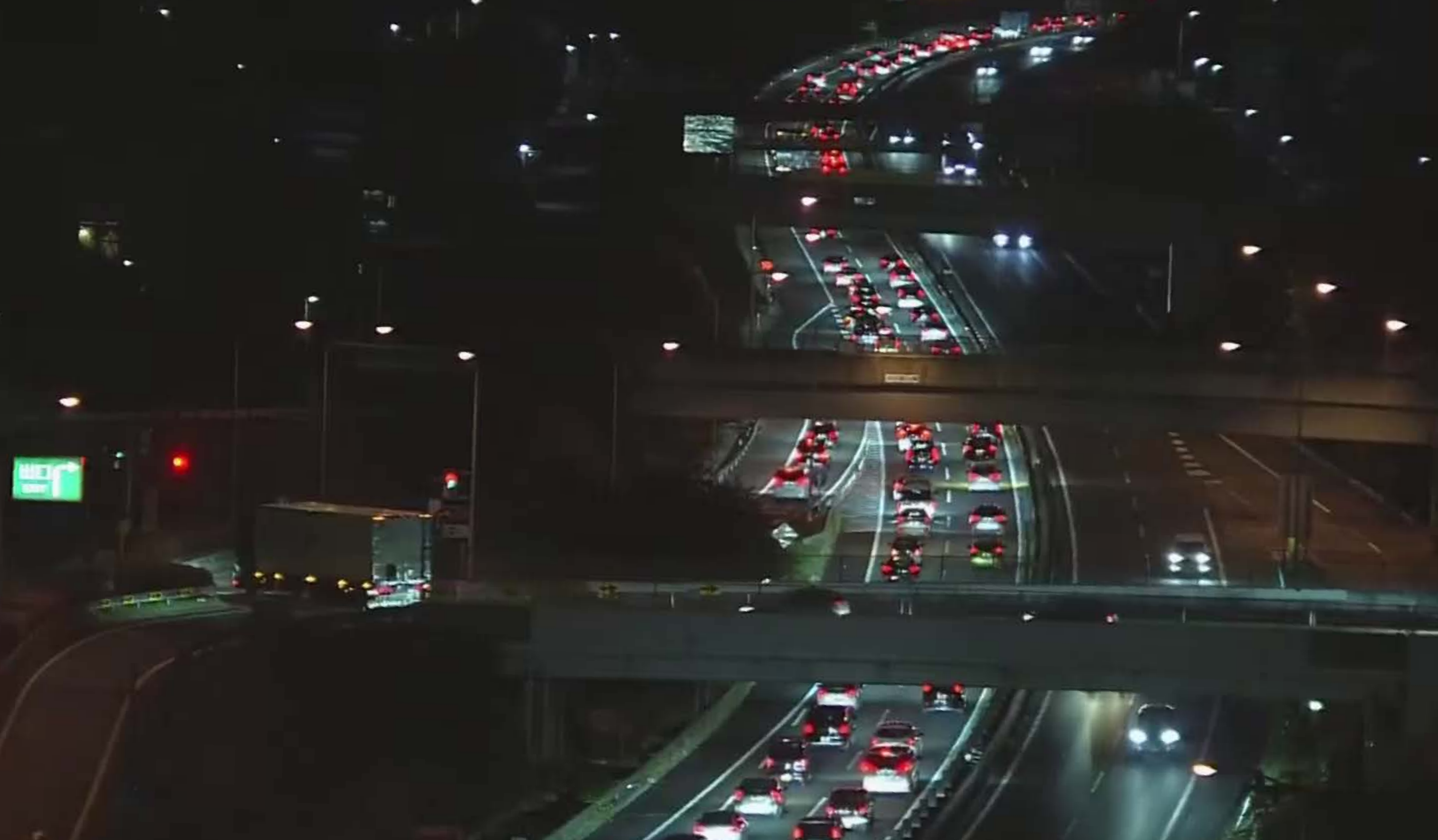}
			\end{minipage}
			\vspace{-5mm}
			\label{Video 2}
		}
		\caption{Illustrations of example frames from two test videos.}
		\vspace{-3mm}
		\label{test frame}
	\end{figure}

\begin{figure}[t]
		\centering
		\includegraphics[width=0.6\linewidth]{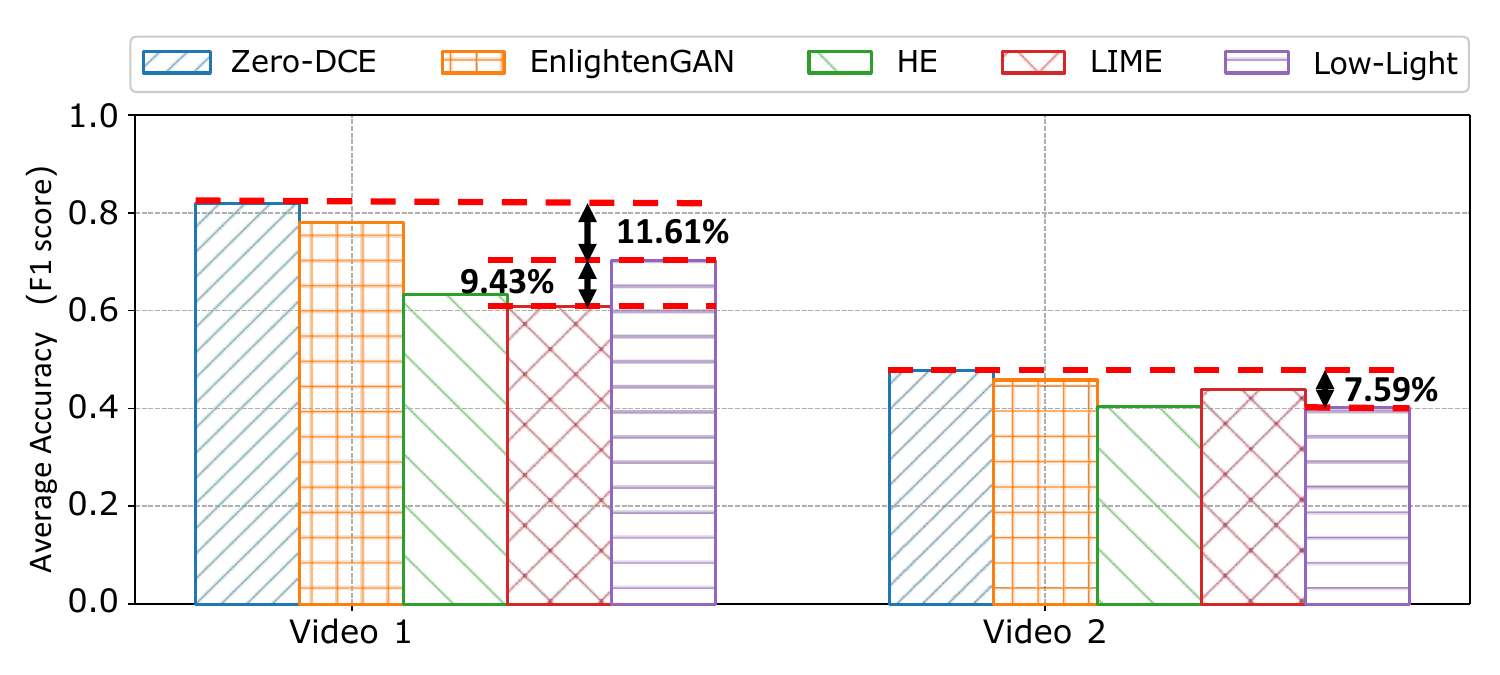}
		\caption{Accuracy of the two videos with or without enhancement.}
		\vspace{-5mm}
		\label{Accuracy of the two videos with or without enhancement}
	\end{figure}

	We explore the impact of different low-light enhancement algorithms on the accuracy of object detection tasks using YOLOv5s\footnote{https://github.com/ultralytics/yolov5. Accessed Apr. 16, 2024.}. Following the enhancement-analytics paradigm\cite{zhangOwlPreandPostprocessing2023, wu2022edge}, we enhance the videos before detection. Four enhancement algorithms are applied, including (1) learning-based algorithms (Zero-DCE\cite{guo2020zero} and EnlightenGAN\cite{jiang2021enlightengan}), (2) theory-based algorithms (HE\cite{coltuc2006exact} and LIME\cite{7782813}). The test videos are derived from YouTube\footnote{
			https://www.youtube.com/watch?v=leskMPI2dN4. Accessed Apr. 16, 2024.
   
			https://www.youtube.com/watch?v=slOgQojt8w8. Accessed Apr. 16, 2024.
    } 
    with a duration of 10 s. The resolution and the frame rate of the videos are 1280$\times$720 and 10 fps, respectively.  As shown in Fig. \ref{test frame}, they exhibit variations in lighting conditions, object distribution, object size, and object velocity of motion. Fig. \ref{Accuracy of the two videos with or without enhancement} illustrates the accuracy evaluated by the F1 score. It shows that for Video 1, the learning-based approaches improve the accuracy up to 11.61\%, while the accuracy of LIME decreases by 9.43\% compared to Low-Light. For Video 2, the accuracy improvement of the enhancement algorithm is limited to 7.59\% at best. This observation underscores the variability in accuracy of low-light videos enhanced by different enhancement algorithms. 
	
	\subsection{Difference Between Enhancement and Low-Light}
\begin{figure}[t]
		\centering
		\subfigure[Unenhanced frame of Video 1]{
			\begin{minipage}[b]{0.45\linewidth}
				\includegraphics[width=1\textwidth]{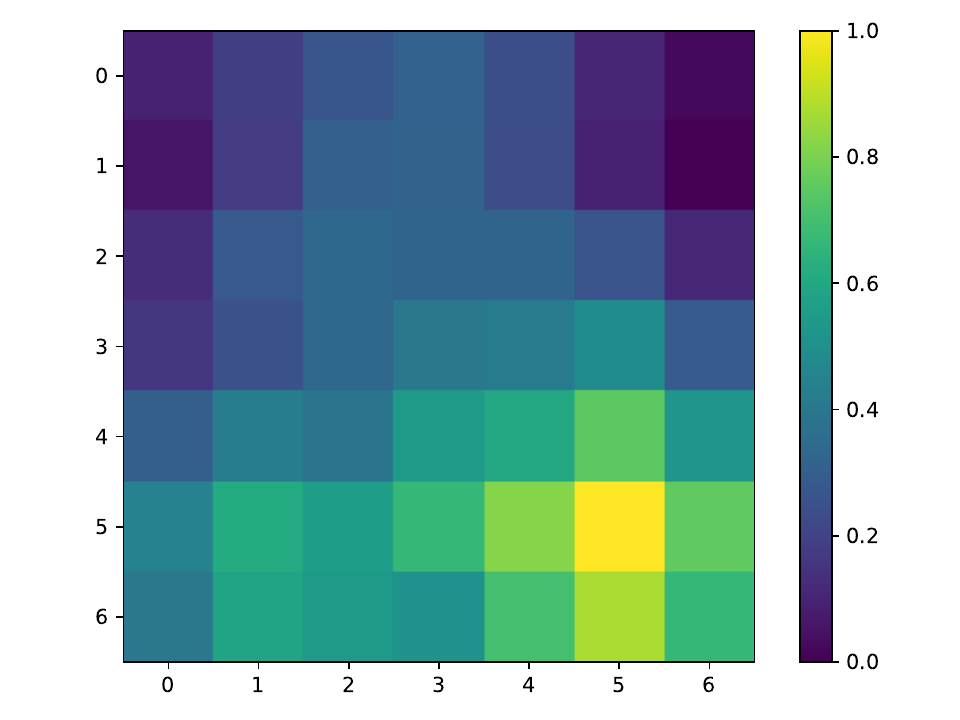} 
			\end{minipage}
			\vspace{-5mm}
			\label{Unenhanced frame of Video 1}
		}
		\subfigure[Unenhanced frame of Video 2]{
			\begin{minipage}[b]{0.45\linewidth}
				\includegraphics[width=1\textwidth]{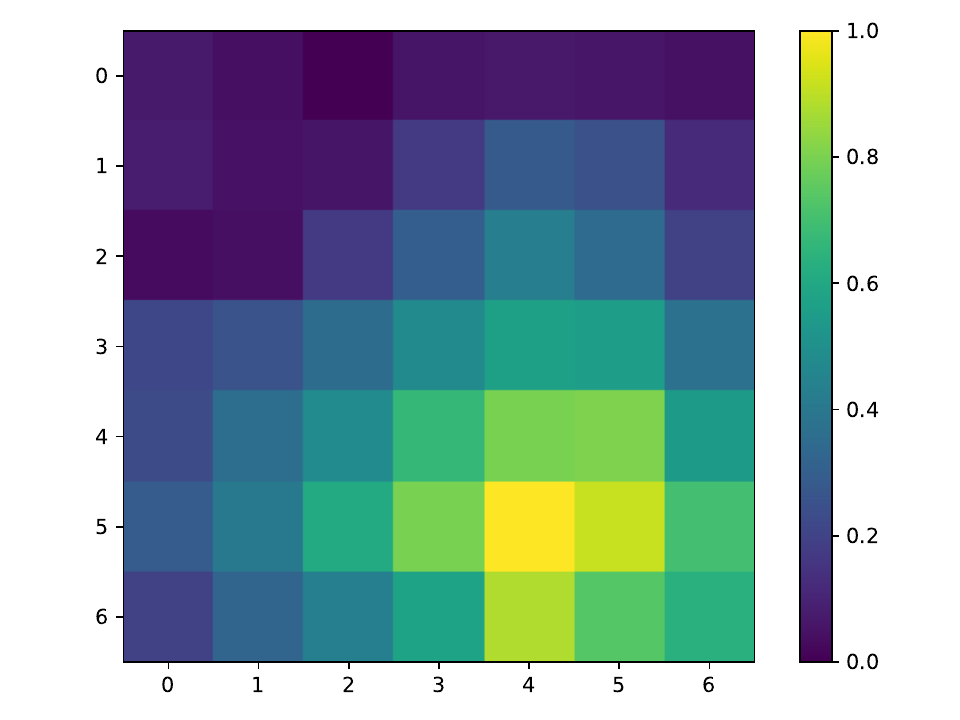}
			\end{minipage}
			\vspace{-5mm}
			\label{Unenhanced frame of Video 2}
		}
		\\ 
		\subfigure[Enhanced frame of Video 1]{
			\begin{minipage}[b]{0.45\linewidth}
				\includegraphics[width=1\textwidth]{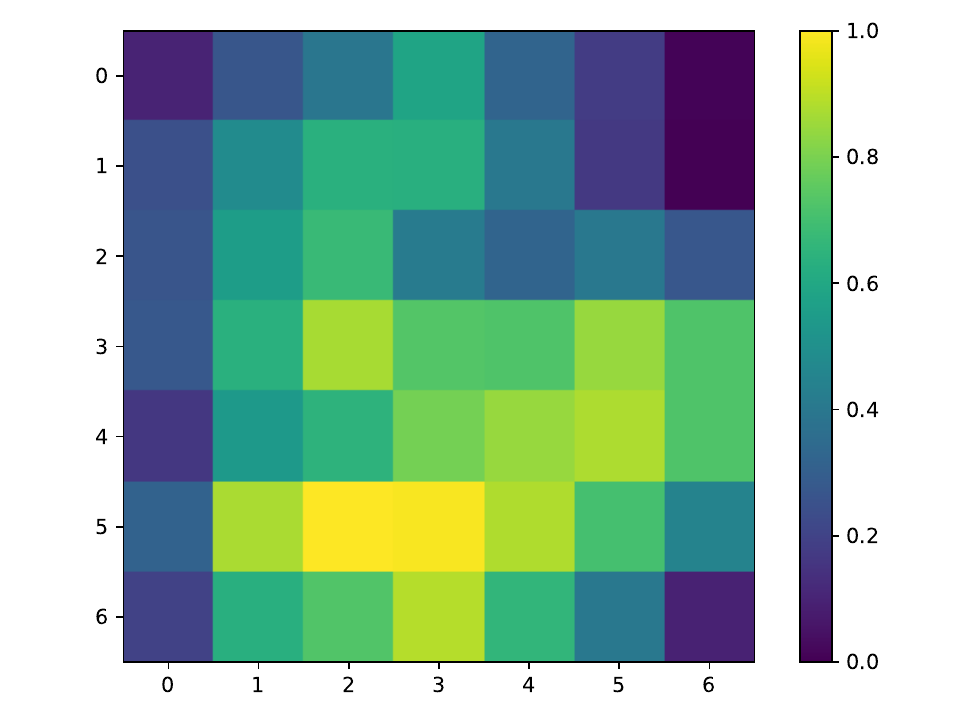} 
			\end{minipage}
			\vspace{-5mm}
			\label{Enhanced frame of Video 1}
		}
		\subfigure[Enhanced frame of Video 2]{
			\begin{minipage}[b]{0.45\linewidth}
				\includegraphics[width=1\textwidth]{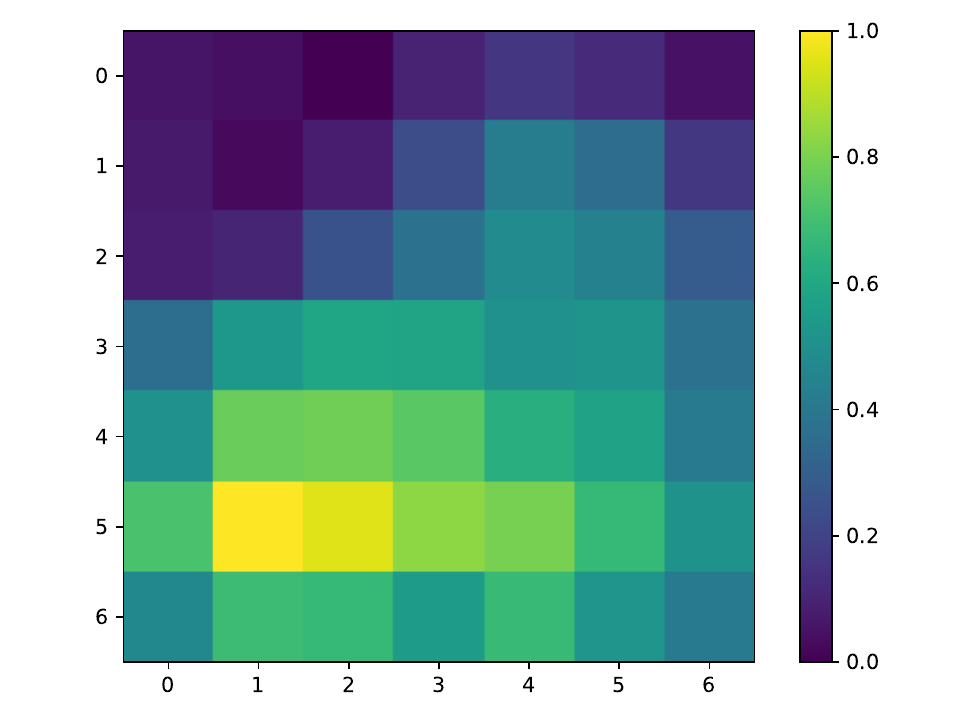}
			\end{minipage}
			\vspace{-5mm}
			\label{Enhanced frame of Video 2}
		}
		\caption{CAM comparisons of different frames.}
		\vspace{-5mm}
		\label{CAM comparisons}
	\end{figure}

\begin{figure}[t]
		\centering
		\includegraphics[width=0.55\linewidth]{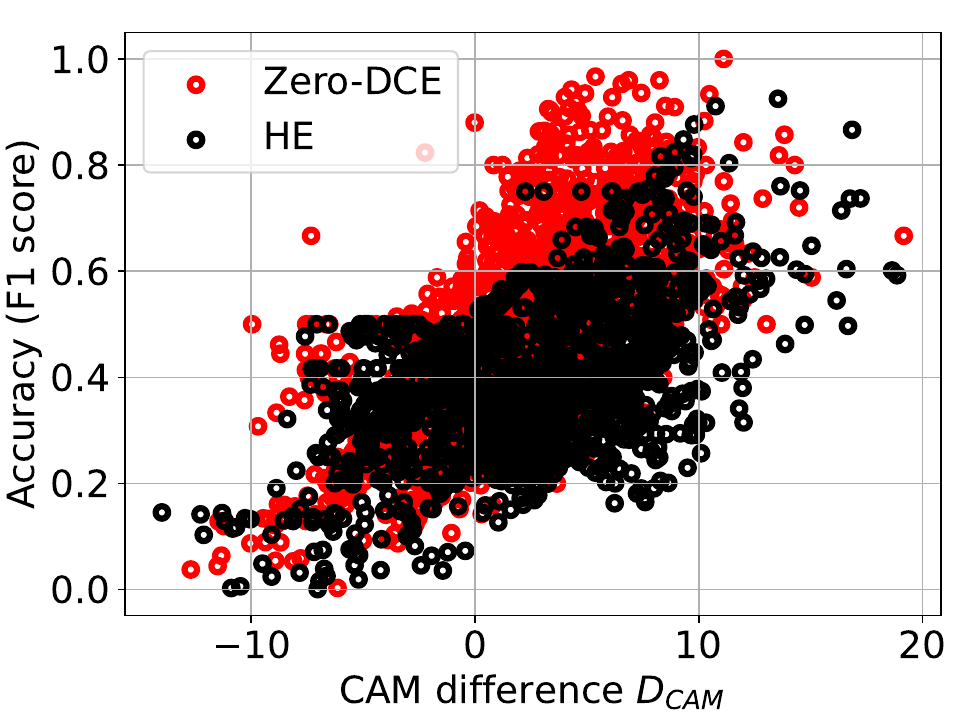}
		\caption{CAM difference between enhanced and low-light frames and the accuracy of the enhanced frames.}
		\vspace{-5mm}
		\label{CAM-diff-vs-Accuracy}
	\end{figure}

	We employ CAMs to explore the object detection model incorporating Convolutional Neural Networks. 
    CAM is derived through a weighted summation of the global average pooling layer and the feature maps from the final convolutional layer\cite{zhou2016learning}. We investigate the CAMs associated with the highest class prediction probability from ResNet18. As depicted in Fig. \ref{Unenhanced frame of Video 1} and Fig. \ref{Enhanced frame of Video 1}, the application of the Zero-DCE to the frame of Video 1 results in an expanded localization range within its CAM, accentuating potential objects. As shown in Figs. \ref{Unenhanced frame of Video 2} and \ref{Enhanced frame of Video 2}, when the frame of Video 2 is also enhanced by Zero-DCE, the region of interest of the CNN model only shifts after enhancement. 
    To evaluate the difference between the enhancement and low-light videos for high-level visual models, We introduce the CAM difference $D_{CAM}$, which is defined as follows
	\begin{equation}
		\label{CAM Difference}
		D_{CAM} = \sum_{i,j}{\left({M}_{E}\left ( i,j \right) - M_{L}\left( i,j \right)\right)},
	\end{equation}
	where ${M}_{E}\left ( i,j \right)$ and $M_{L}\left( i,j \right)$ are the values of the $i$-th row and the $j$-th column of the CAM of the enhanced frame and the low-light frame, respectively. We extend the duration of Video 1 to 100 s and apply Zero-DCE and HE for enhancement. As shown in Fig. \ref{CAM-diff-vs-Accuracy}, the CAM difference between enhanced and low-light frames and the accuracy of the enhanced frames exhibits a positive correlation, i.e., frames with larger $D_{CAM}$ tend to have higher accuracy. Furthermore, the frames enhanced with Zero-DCE generally exhibit higher accuracy compared to those enhanced with HE, particularly when $D_{CAM} > -2.5$. It indicates that the difference between the enhanced and low-light frames of Video 1 is relatively large, and Zero-DCE takes advantage of this difference to provide more significant enhancement effects.
 
	Through the above analysis, it can be inferred that the effectiveness of enhancement algorithms in improving the accuracy is partially contingent upon the specific video. Enhancing low-light videos does not consistently improve analytics accuracy. Furthermore, CAM provides the opportunity to contrast the difference between enhanced and low-light frames, making it possible to assess the effectiveness of enhancement.
	
\section{System Model and Problem Formulation}
	\subsection{System Overview}
\begin{figure}[t]
		\centering
		\includegraphics[width=0.9\linewidth]{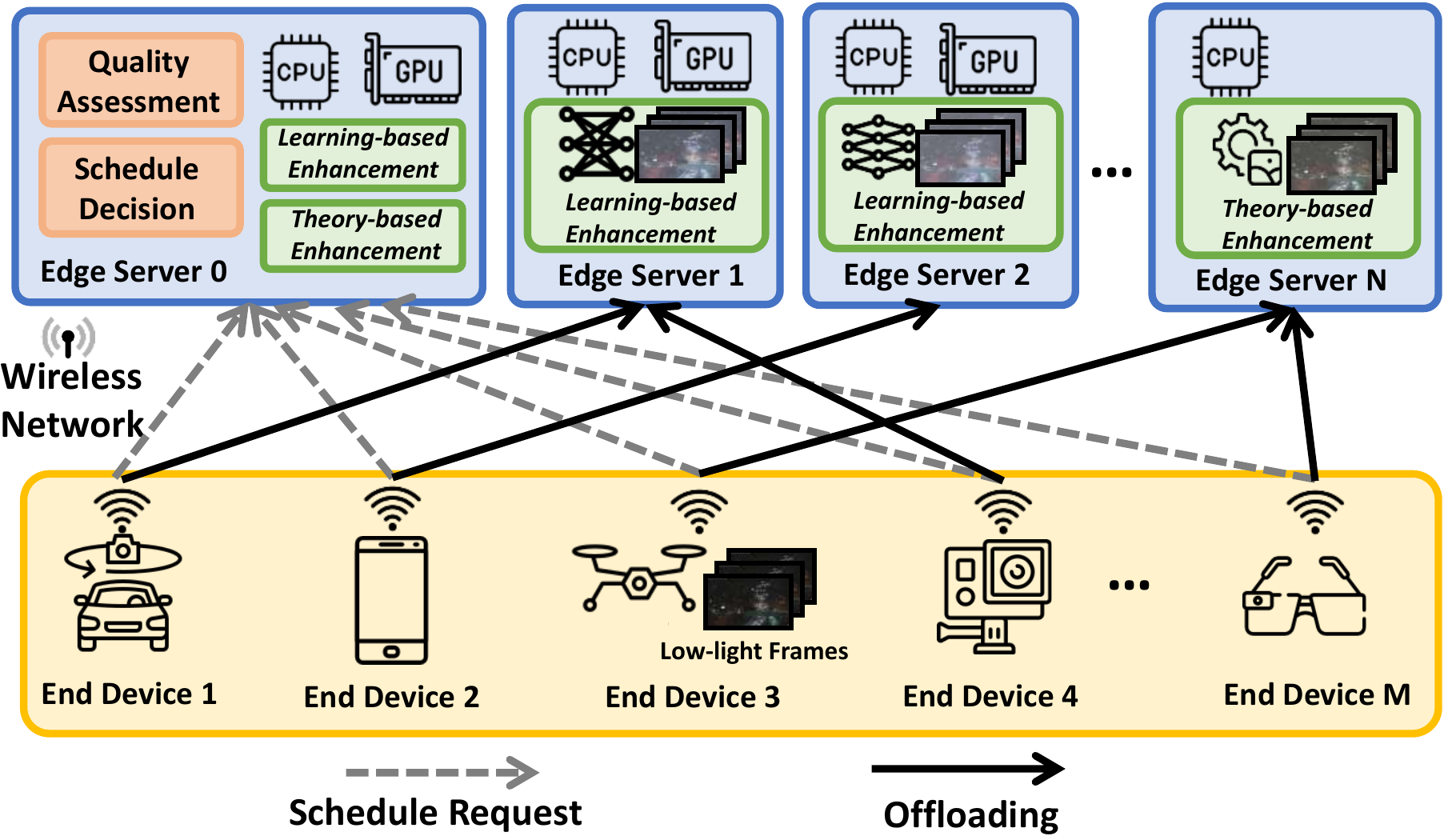}
		\caption{System Model}
		\vspace{-5mm}
		\label{System Model}
	\end{figure}

    In our system, there are $N + 1$ geographically distributed edge servers providing low-light video analytics services through a wireless network to $M$ end devices, in the set of $\mathcal{M}=\{ 1,2,\cdots, m, \cdots, M\}$. Together they form an edge cluster $\mathcal{N}$ where $\mathcal{N}=\{ 0,1,2,\cdots, n, \cdots ,N\}$. As shown in Fig. \ref{System Model}, edge servers have different computation resources for running the enhancement algorithms, and the computation capacity of the $n$-th edge server is defined as $C^{n}$. There are $K$ enhancement algorithms, including learning-based and theory-based algorithms, which form the set $\mathcal{K}=\{ 1,2,\cdots, k, \cdots, K\}$. 
    To fully leverage the end-edge bandwidth resources and computation resources of edge servers, the $0$-th edge server is treated as the schedule controller, which receives scheduling requests from end devices and responds to them with scheduling decisions for enhancement and analytics. 
    It should be equipped with adequate computation resources to run all the enhancement algorithms. Additionally, it needs the capability to rapidly calculate the CAMs for each low-light frame and the corresponding enhanced frame, ensuring the computation demands for the enhancement assessment. Moreover, the $0$-th edge server must establish stable and low-latency communication with all end devices to ensure timely receipt and transmission of scheduling requests and decisions.
    After receiving the schedule information, end devices offload the low-light video tasks to the scheduled edge server. We use a binary variable $u_{t}^{m,n}$ to indicate whether the scheduling controller selects the $n$-th server in time slot $t$ for offloading the task of the $m$-th end device. At the same time, the binary variable $x_{t}^{m,k}$ is denoted to the choice of the $k$-th enhancement algorithm for videos from the $m$-th end device, and $x_{t}^{m,0}$ means no enhancement. 

	\subsection{Enhancement Quality Assessment}
	For a low-light video, it is divided into chunks of the same duration. To accommodate real-time video analytics, a chunk is not too long, where we assume that the video content will not change much, and the first frame in a chunk is selected to generate a CAM. In each time slot, the end device sends the encoded first frame to the schedule controller. Then, all the enhancement algorithms are used on this low-light frame to obtain the enhanced CAMs, where $M_{E,t}^{m,k}$ is the CAM generated from the enhanced frame sent by the $m$-th end devices using the $k$-th enhancement algorithm. In addition, the low-light CAM of the frame $M_{L,t}^{m,k}$ is also extracted. After that, to concentrate the quality assessment on areas where objects might be located, regions of a CAM exceeding the threshold $\gamma$ are preserved as follows
	\begin{equation}
		\label{filter CAM}
		\tilde{M}_{E,t}^{m,k}=\left\{ \left( i,j \right) \in M_{E,t}^{m,k}\left| M_{E,t}^{m,k}\left( i,j \right) >\gamma \right. \right\},
	\end{equation}
	where, $\tilde{M}_{E,t}^{m,k}$ denotes the filtered CAM composed of elements greater than the threshold $\gamma$ in $M_{E,t}^{m,k}$, and $M_{E,t}^{m,k}\left( i,j \right)$ represents the element in the $i$-th row and $j$-th column of $M_{E,t}^{m,k}$. 
	
	Then, the enhancement quality can be defined as follows
	\begin{equation}
		\label{enhancement quality}
		\begin{split}
			Q_{t}^{m,k} = \frac{\alpha _t^m \sum_{i,j}{\left( \tilde{M}_{E,t}^{m,k}\left( i,j \right) - \tilde{M}_{L,t}^{m,k}\left( i,j \right)\right) }}{\sum_{w=1}^W{\sum_{i,j}{\left| \tilde{M}_{E,t}^{m,k}\left( i,j \right) -\tilde{M}_{E,t-w}^{m,k}\left( i,j \right) \right|}}},
		\end{split}
	\end{equation}
	where $\alpha_t^m$ is the average accuracy of the last $W$ chunks of the $m$-th end devices, used to regulate the weight in $Q_{t}^{m,k}$. And $Q_{t}^{m,0} = 0$ indicates no enhancement does not have an impact on analytics accuracy. The numerator indicates the sum of the differences between $\tilde{M}_{E,t}^{m,k}$ and $\tilde{M}_{L,t}^{m,k}$, where $\tilde{M}_{L,t}^{m,k}$ is the filtered low-light CAM of $M_{L,t}^{m,k}$, analogous to Equation (\ref{filter CAM}). It reveals whether and how much the $k$-th enhancement algorithm is effective in improving accuracy. The denominator indicates the sum of the differences between the current filtered CAM $\tilde{M}_{E,t}^{m,k}$ and the filtered CAMs of the past $W$ chunks. Empirically, due to the continuity of video content, the analytics results also remain stable. Therefore, the current chunk is likely to hold similar results to previous chunks. 
	
	\subsection{Problem Formulation}
	Once the enhancement quality is assessed, the schedule controller needs to estimate the latency in time slot $t$, which mainly comes from the transmission latency from end devices to edge servers, and the enhancement latency on edge servers. Without loss of generality, schedule latency, and inference latency are considered as constant $L_O$ in our system\footnote{The schedule latency comprises the latency of enhancement quality assessment and that of the scheduling algorithm. The latency of enhancement quality assessment can be recorded by the schedule controller. The latency of the scheduling algorithm depends on its complexity and the computation resources at the schedule controller. The inference latency can be estimated based on the historical data of the inferred chunks. Since the datasize of the encoded first frame, schedule information, and the detection result are only a few kilobytes, the latency of transmitting them can be ignored. Therefore, the schedule latency and inference latency are regarded as constants.}. Thus, the latency $L_t^m$ of response to the $m$-th end device can be formulated as follows
	\begin{equation}
            \label{latency}	L_t^m=\sum_{n=1}^N{u_{t}^{m,n}l_{T,t}^{m,n}}+\sum_{n=1}^N{u_{t}^{m,n}\sum_{k=0}^K{x_{t}^{m,k}l_{E,t}^{m,n,k}}}+L_O,
	\end{equation}
	where $l_{T,t}^{m,n}$ is the transmission latency of transmitting the video from the $m$-th end device to the $n$-th edge server. It can be denoted as 
        \begin{equation}
            \label{transmission latency}	
            l_{T,t}^{m,n} = \frac{d_{t}^{m}}{b_{t}^{m,n}},
        \end{equation} 
    where $d_{t}^{m}$ is the datasize of the video sent by the $m$-th end device, and $b_{t}^{m,n}$ is the allocated bandwidth between the $m$-th end devices and the $n$-th edge server. $l_{E,t}^{m,n,k}$ is the enhancement latency of the video sent by the $m$-th end device with the $k$-th enhancement algorithm on the $n$-th edge server in time slot $t$, which can be denoted as 
        \begin{equation}
            \label{enhance latency}	
            l_{E,t}^{m,n,k} = \frac{\varphi ^{n,k} d_{t}^{m}}{S^{n,k}},
        \end{equation} 
    where $\varphi ^{n,k}$ is the scaling factor that quantifies the computation demand per bit in terms of GPU floating-point operations (FLOPs/bit) of learning-based enhancement algorithms and CPU cycles (cycles/bit) of theory-based enhancement algorithms. And $S^{n,k}$ is the computation workload of the $n$-th edge server to execute the $k$-th enhancement algorithm (FLOPS or cycles/s). And $l_{E,t}^{m,n,0} = 0$ means no enhancement brings no enhancement latency. 
    
    For real-time low-light video analytics, the enhancement quality and the latency are the key factors in overall system performance. Therefore, the utility function of the $m$-th end device requires a trade-off between the enhancement quality and the latency, which can be formulated as follows
	\begin{equation}
		\label{utility function}
		U_t^m=\sum_{k=0}^K{x_{t}^{m,k}Q_{t}^{m,k}-\varsigma L_t^m},
	\end{equation}
	where $\varsigma$ is the weighting factor between enhancement quality and latency, and $Q_{t}^{m,k}$ is the enhancement quality of the video sent by the $m$-th end device and enhanced with the $k$-th enhancement algorithm in time slot $t$. 
	
	Above all, the schedule controller selects appropriate offloading strategies and enhancement algorithms based on the requested videos of the end devices, the computation resources of the edge servers in the edge cluster, and the latency. For this purpose, we establish the following optimization problem
	\begin{equation}
		\label{minPC}
		\begin{split}
			&\mathcal{P}_0:\ \ \max_{x_{t}^{m,k},u_{t}^{m,n}}\ \sum_{m=1}^M{U_t^m}, \\
			\text{s.t.}\ \ &C_1:\ x_{t}^{m,k}\in \left\{ 0,1 \right\} ,\ k\in \left\{ 0 \right\} \cup \mathcal{K},\ m\in \mathcal{M},\\
			&C_2:\ u_{t}^{m,n}\in \left\{ 0,1 \right\} ,\ n\in \mathcal{N} \setminus \{0\},\ m\in \mathcal{M}, \\
			&C_3:\ \sum_{k=0}^K{x_{t}^{m,k}=1}, \sum_{n=1}^N{u_{t}^{m,n}=1}, \ m\in \mathcal{M},\\
			&C_4:\ \sum_{m=1}^M{u_{t}^{m,n}\sum_{k=0}^K{x_{t}^{m,k}S^{n,k}}}\le C^{n},\ n\in \mathcal{N} \setminus \{0\},\\
			&C_5:\ L_t^m\le L_{max}, \ m\in \mathcal{M},\\
			&C_6:\ t\in \left[ 0,T \right] ,\ t\in \boldsymbol{Z},
		\end{split}
	\end{equation}
	where $L_{max}$ is the maximum latency in time slot $t$. Constraints $C_1$ and $C_2$ are the ranges of $x_{t}^{m,k}$ and $u_{t}^{m,n}$. Constraint $C_3$ indicates that a chunk is only processed and analyzed by one edge server and only one enhancement algorithm can be selected in time slot $t$. In addition, constraint $C_4$ shows a limit on the computation workloads of all edge servers, providing them some stability and avoiding computation crashes. Constraint $C_5$ ensures the latency does not exceed the latency upper bound $L_{max}$. Constraint $C_6$ is the range of time slots.
\subsection{Algorithm Design}
\begin{algorithm}[t]
    \caption{Genetic-based scheduling.}\label{alg1}
    \begin{algorithmic}[1]
    \State Randomly generate $V$ individuals as the initial population, set generation start index $i = 1$
    \For {each generation  $i = 1$ to $I$}
    \State \textbf{Fitness:} Calculate the utility of each individual $\boldsymbol{O}$ 
	\State \parbox[t]{0.85\linewidth}{\raggedright Retain the individual with the highest utility to the next generation}
	\For {$v = 1$ to $V-1$}
	\State \parbox[t]{0.85\linewidth}{\raggedright \textbf{Selection:}  Randomly select two individuals $\boldsymbol{O}_1$ and $\boldsymbol{O}_2$ from the population based on the utility of each individual.}
	\State $\boldsymbol{O}' = \boldsymbol{O}_1$
	\State \parbox[t]{0.85\linewidth}{\raggedright \textbf{Crossover:} Conduct $\boldsymbol{O}' = Crossover(\boldsymbol{O}_1, \boldsymbol{O}_2)$ with probability $P_c$}
	\State \parbox[t]{0.85\linewidth}{\raggedright \textbf{Mutation:} Conduct $\boldsymbol{O}' = Mutation(\boldsymbol{O}')$ with probability $P_m$}
        \EndFor
	\EndFor
    \end{algorithmic}
\end{algorithm}

Let $\boldsymbol{\chi }$ and $\boldsymbol{\mu }$ be the decision vectors of the optimization problem $\mathcal{P}_0$, which is a 0-1 integer programming problem. To find the optimal solution, one needs to evaluate the objective function for each solution, and the complexity of this process is $O(2^{M (N + K)})$. However, such a brute-force search approach is computationally expensive and time-consuming. Timely responses are crucial to our system, it cannot afford exhaustive search of every possible solution. Therefore, we need to find a near-optimal solution reasonably to meet the latency requirement.
The algorithm is outlined in Algorithm \ref{alg1}. For expression simplicity, we denote the decision pair $(\boldsymbol{\chi}, \boldsymbol{\mu})$ as an individual by $\boldsymbol{O}$. For the initialization process, we randomly generate $V$ individuals. Then, at each generation, we calculate the utility of each individual and retain the individual with the highest utility. The remaining individuals are further processed (line 6-9) to create the next generation.  In the $Corssover$ operation, a new individual $\boldsymbol{O}'$ is generated by inheriting parts of the genes of the parent individuals ($\boldsymbol{O}_1$ and $\boldsymbol{O}_2$). Meanwhile, the $Mutation$ operation randomly alters the gene at a random position. A total of $I$ iterations are conducted to gradually improve the quality of each individual. Finally, we obtain the problem solution by selecting the best individual in the last generation. With the proposed genetic-based solution, the problem is solved at a complexity of $O(VI)$, which guarantees efficient online scheduling.
\section{Performance Evaluation}
\subsection{Experimental Settings}
\begin{table}[t]
	\centering
    	\caption{Enhancement Algorithms}
	\begin{tabular}{ c | c | c } 
		\hline\hline
		  \textbf{Enhancement Algorithm} & \textbf{Platform}  & \textbf{Type}\\  
		\hline\hline
		HE\cite{coltuc2006exact} & OpenCV(CPU) & Theory-based  \\
		\hline
		LIME\cite{7782813} & MATLAB (CPU) & Theory-based  \\
		\hline
            Zero-DCE\cite{guo2020zero} & PyTorch (GPU) & Learning-based   \\ 
		\hline
		EnlightenGAN\cite{jiang2021enlightengan} & PyTorch (GPU) & Learning-based  \\
		\hline
	\end{tabular}
	\vspace{-5mm}
	\label{table:Algorithms}
\end{table}
\textbf{Platforms.} All experiments are conducted on a testbed equipped with Intel(R) Core(TM) i9-11900K @ 3.50 GHz CPU and NVIDIA GeForce 3080Ti GPU. The video analytics task involves object detection using YOLOv5s. CAM is generated using the lightweight feature extractor ResNet18. The serial number and computation capacity of edge servers is set to \{1: 34.1 TFLOPS GPU and 3.5 GHz CPU, 2: 1.5 TFLOPS GPU and 854 MHz CPU, 3: 2 GHz CPU, 4: 1 GHz CPU\}. The enhancement algorithms are shown in Table \ref{table:Algorithms}. The set of hyper-parameters \{$M$, $N$, $K$, $\varsigma$, $L_{max}$, $W$, $b_t^m$\} is set to \{10, 4, 4, 0.5, 4 s, 5, 20 Mbps\} by default.

\textbf{Dataset.} Apart from the two videos in Section \ref{Detection of different low-light videos}, we add another two experimental videos from datasets, \cite{che2019d} and \cite{cao2021visdrone}. We employ the FFmpeg\footnote{https://ffmpeg.org/. Accessed Apr. 16, 2024.} to encode the videos. These videos encompass a range of low-light scenarios, including weak blurred illumination, point light sources, and shadows. The frames are encoded with a resolution of 1280$\times$720, a frame rate of 10 fps, and a QP of 22. Each experimental video has a duration of 30 s, with chunks of 1 s.

\textbf{Baseline.} We conduct comparisons between our proposed system and several baseline methods. 
\begin{itemize}
	\item \textbf{Capacity-based:} This method selects an enhancement algorithm based solely on whether the computation workload of the end devices matches the capacity of the edge server. Tasks that exceed the computation capacity of edge servers will be rejected.
	\item \textbf{Fixed-End-Edge\cite{he2023globecom}:} The end device consistently offloads all enhancement tasks to a fixed-edge server without adaptability. It streams chunks with different QP according to bandwidth resources, while the edge server performs enhancement with one enhancement algorithm. 
	\item \textbf{JCAB-Muti\cite{9155524}:} It is a state-of-the-art frame-based method that optimizes both encoding configuration adaptation and bandwidth allocation in multi-edge-based video analytics systems.
\end{itemize}
\subsection{Performance Comparison}
    \begin{figure}[t]
	\centering
	\subfigure[Average accuracy]
	{
		\begin{minipage}[b]{.44\linewidth}
			\centering
			\includegraphics[width=\textwidth]{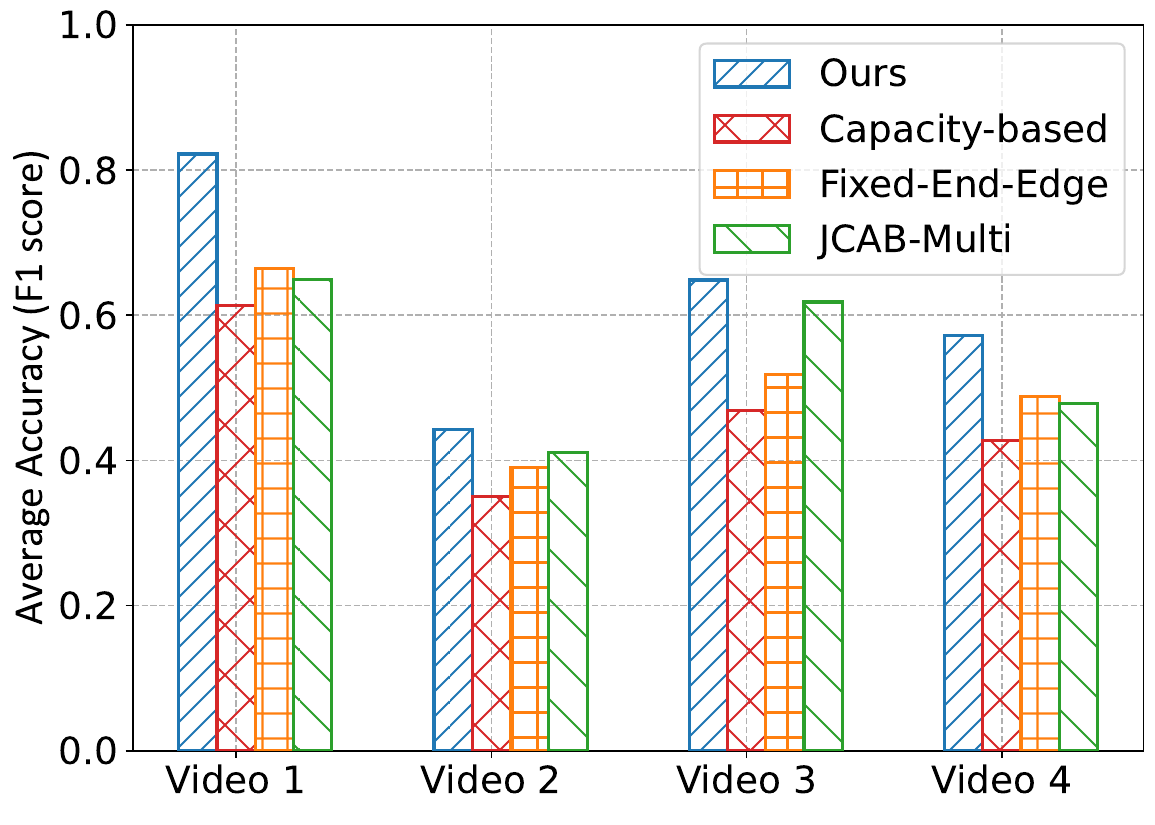}
			\vspace{-5mm}
			\label{Accuracy-dm}
		\end{minipage}
	}
	\subfigure[Average latency]
	{
		\begin{minipage}[b]{.44\linewidth}
			\centering
			\includegraphics[width=\textwidth]{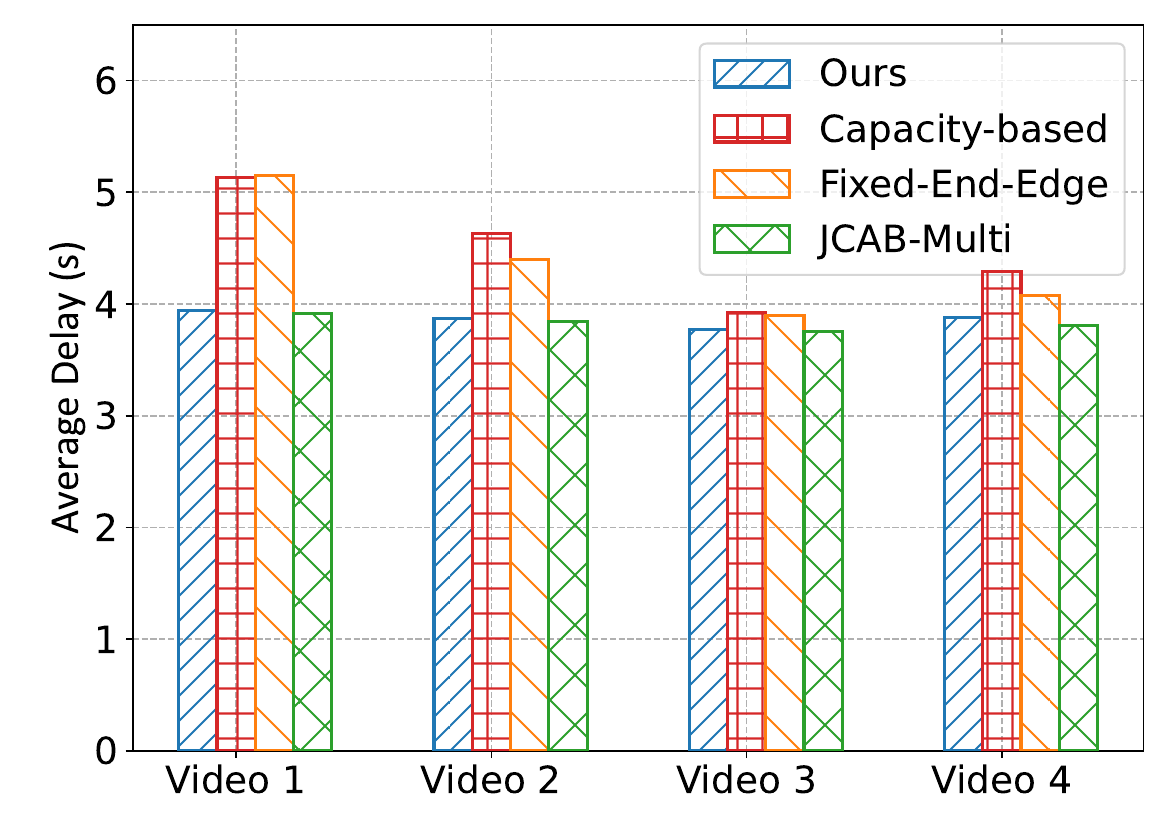}
			\vspace{-5mm}
			\label{Latency-dm}
		\end{minipage}
	}
	\caption{Performance comparison on average accuracy and latency.}
	\vspace{-5mm}
	\label{fig: ev_accu_delay}
    \end{figure}
    For the videos with diverse contents, we investigate the inference accuracy and latency performance across different methods. As depicted in Fig. \ref{Accuracy-dm}, our proposed system demonstrates superior accuracy compared to alternative methods. Specifically, we observe an improvement in accuracy by up to 20.83\% in comparison to the Capacity-based. And it brings up to a 15.76\% accuracy increase compared to Fixed-End-Edge. This underscores the significance of multi-edge server scheduling in enhancing low-light videos. Furthermore, our approach surpasses JCAB-Multi with an accuracy improvement ranging from 3.03\% to 17.33\% across different videos. Despite the effectiveness of JCAB-Multi in encoding configuration and bandwidth allocation, it overlooks the potential for accuracy improvement through enhancement as it does not cater to low-light videos.
    In regards to latency, our proposed method exhibits lower latency. As illustrated in Fig. \ref{Latency-dm}, when compared to JCAB-Multi without enhancement, the latency only experiences a slight increase ranging from 14.14 ms to 72.30 ms across different videos due to the enhancement processing, while achieving higher accuracy. Moreover, the Capacity-based approach, which solely considers edge server computation resources, fails to meet the latency requirements of end devices. Although adapting the enhancement model, the Fixed-end-edge approach struggles to accommodate multi-end requests, resulting in longer latency.
\subsection{Impact of System Parameters}
    \begin{figure}[t]
	\centering
	\subfigure[The impact of bandwidth on accuracy]
	{
		\begin{minipage}[b]{.45\linewidth}
			\centering
			\includegraphics[width=\textwidth]{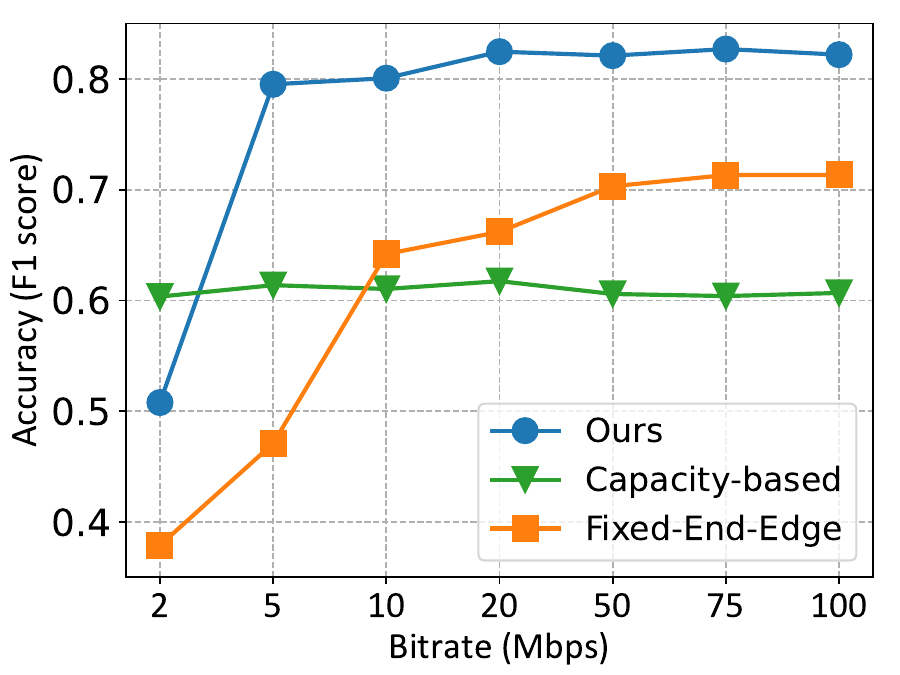}
			\vspace{-5mm}
			\label{Accuracy-bw}
		\end{minipage}
	}
	\subfigure[The impact of end devices on accuracy]
	{
		\begin{minipage}[b]{.45\linewidth}
			\centering
			\includegraphics[width=\textwidth]{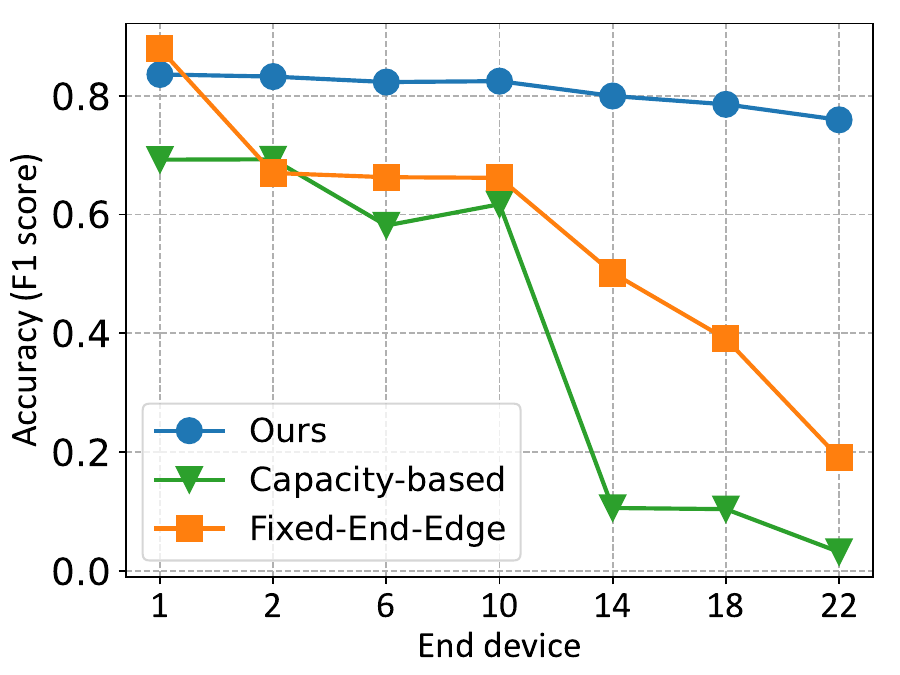}
			\vspace{-5mm}
			\label{Latency-end}
		\end{minipage}
	}
	\caption{Performance comparison across various system parameters.}
	\vspace{-5mm}
	\label{fig: ev_bw_la}
    \end{figure}
    \subsubsection{The impact of bandwidth variation on accuracy}
    We uniformly allocate the same bandwidth between each end device and the edge server to simplify the design. The bandwidth directly affects transmission latency. In Fig. \ref{Accuracy-bw}, we compare the accuracy of Video 1 for our proposed method with the Capability-based and Fixed-End-Edge methods under situations of different bandwidth. The accuracy of our method decreases when the bandwidth is less than 20 Mbps. This reduction is attributed to the increase in latency, which leads our method to prioritize reducing the enhancement while ensuring latency requirements. Notably, our method outperforms the Fixed-End-Edge approach in accuracy for all bandwidth conditions. The Capability-based method, however, does not consider bandwidth variations, resulting in stable accuracy but an inability to meet latency requirements, e.g., the latency is 8.13 s when the bandwidth is 2 Mbps.
    \subsubsection{The impact of the number of end devices on accuracy}
    We adjust the number of end devices to explore its impact on accuracy. As shown in Fig. \ref{Latency-end}, when there are 22 end devices, our method exhibits a slight decrease in accuracy by 7.64\% compared to the scenario with 2 end devices. Conversely, both the Capability-based and Fixed-End-Edge methods experience significant drops in accuracy when the number of end devices exceeds 10. When the computation capacity of the edge servers is exceeded, tasks offloaded from the end devices are rejected and discarded, which results in a severe drop in the accuracy of the Capacity-based method. Although Fixed-End-Edge is adaptive in terms of end-edge coordinated encoding and enhancement, it is constrained to the scenario of fixed end-edge configurations.
    \section{Conclusion}
    In this paper, we have designed a multi-edge system, adaptively offloading and enhancement for low-light video analytics. Differences in analytics accuracy have been observed across low-light videos processed with different enhancement algorithms. We have introduced a novel enhancement quality assessment method based on CAM differences between enhanced and low-light frames. Then, a multi-edge offloading and enhancement system has been proposed under limited end-edge bandwidth and edge server computation resources. Simulation experiments demonstrate the superiority of our proposed system in improving accuracy and reducing latency compared to the baselines. In future work, we will investigate machine vision characteristics and video analytics tasks in adverse environments.

    \section*{Acknowledgment}
    The work was supported in part by the Natural Science Foundation of China under Grant 62001180, and in part by the Young Elite Scientists Sponsorship Program by CAST under Grant 2022QNRC001.


\bibliographystyle{IEEEtran}	
\bibliography{ref}

\end{document}